\documentclass{aa}
\usepackage{graphicx}
\begin{document}
%

\title{RXTE observations of single pulses of PSR B0531+21: II Test for 
       radio behavior}

\titlerunning{RXTE observations of single pulses of PSR B0531+21}

\author{M. Vivekanand\thanks{vivek@ncra.tifr.res.in}}

\institute{National Center for Radio Astrophysics, TIFR, Pune University Campus,
P. O. Box 3, Ganeshkhind, Pune 411007, India.}

\date{Received (date) / Accepted (date)}
\abstract{
This article is the second in the series that analyze about 1.87 million periods 
of Crab pulsar, observed by the PCA detector aboard the RXTE x-ray observatory.
At these energies the pulsar displays none of the three phenomena that are often 
seen in normal radio pulsars -- ``pulse nulling'', ``systematic sub pulse drifting'' 
and ``mode changing''. Presence or absence of these three behavior in the Crab
pulsar, at radio wavelengths, something that has not been rigorously established 
yet, might be important for a satisfactory understanding of the above three 
phenomena.
}
\maketitle

\section{Introduction}

The coherent microwave radio emission mechanism of rotation powered pulsars 
is as yet an unsolved problem. The Ruderman and Sutherland \cite{RS} model 
is the most successful in explaining some of the observed radio behavior. 
It postulates a region close to the neutron star surface that accelerates 
charges to relativistic energies, which produces $\gamma$ ray photons by 
synchrotron or related mechanism. These $\gamma$ rays produce e$^+$--\,e$^-$
pairs in strong magnetic fields. The pairs are accelerated to 
relativistic energies in the same accelerator region. This process cascades 
into a spark of e$^+$--\,e$^-$ pair plasma, which eventually produces the 
observed coherent microwave radio radiation. However the model requires a 
high work function of iron ions on the surface of the neutron star, which was 
later calculated to be too low (Hillebrandt \& Mueller \cite{HM76}; Flowers 
et al. \cite{FRHM77}). Cheng and Ruderman (\cite{CR}, and references therein) 
salvaged the model by invoking e$^+$--\,e$^-$ pair production by thermal 
x-ray photons, emitted by the hot neutron star surface, in the Coulomb field 
of relativistic iron ions that are no longer bound to the neutron star 
surface. Therefore thermal x-ray photons probably play a significant role in 
the pulsar coherent radio emission mechanism. 

The above models are classified as {\it inner-gap} (or polar cap) models, since 
their accelerator is just above the polar cap surface of the neutron star. 
However, the higher energy radiation from these pulsars ($\gamma$ rays, x-rays, 
infra-red and optical) is now believed to originate in the {\it outer gaps} 
(Cheng et al. \cite{CHR1}, \cite{CHR2}, henceforth CHR; Romani \& Yadigaroglu 
\cite{RY95}), in which the accelerator is far away from the neutron star 
surface. This would have been a neat division of labor between the two gaps, 
but for the fact that the integrated profile of Crab pulsar is aligned (i.e., 
arrives at the same time), and also looks similar in shape, at wavelengths 
ranging from radio to $\gamma$ rays, except for a small part known as the radio 
precursor (Smith, \cite{FGS86}; Lundgren et al. \cite{LCM95}; Moffet \& Hankins 
\cite{MH1996}). It is now speculated that most of the Crab pulsar emission 
(including the radio) arises in the outer gap, while the radio precursor arises 
in the inner gap (CHR). In the outer gap models, the observed x-rays are 
produced by synchrotron processes by the {\it secondary} e$^+$--\,e$^-$ pairs, 
which also play a role in the {\it bootstrap} functioning of the outer gap 
(CHR). Thus at least in the Crab pulsar, there appears to be an intimate 
connection between the coherent radio emission and the non-thermal x-ray 
emission also (although this may not be true for, say, the Vela pulsar).

This paper looks for, in the Crab pulsar at x-ray energies, the three phenomena 
that are commonly observed at radio wavelengths in several normal (i.e., non 
milli second) rotation powered pulsars -- pulse ``nulling'', systematic subpulse 
``drifting'', and integrated profile instability or ``mode changing''. The 
motivation for this analysis is to establish the direct connection, if any, 
between the radio and x-ray emitting relativistic charges in rotation powered
pulsars.

Consider pulse nulling; it is as yet an unexplained phenomenon. Filippenko \& 
Radhakrishnan (\cite{FR1982}) argue that the absence of radio emission during 
nulling is due to the cessation of sparking, which leads to a cessation of 
bunching of charges, and consequently to loss of coherence. However the current
of relativistic charges exists, flowing continuously rather than as a series of 
sparks. Now, it is generally believed that the radio emission of rotation 
powered pulsars is due to a coherent process while the higher energy emissions 
(optical, x-ray, $\gamma$-ray, etc) are due to incoherent processes (see 
Manchester \& Taylor \cite{MT1977}). Therefore in the above picture, radio 
nulling need not be accompanied by nulling of the high energy emission. On the 
other hand, Kazbegi et al. (\cite{KMMS1996}) think that pulse nulling is caused 
by ``very low-frequency drift waves'' that ``change the curvature 
radius of field lines''. Since a fundamental parameter is affected here, it is 
likely that in this scenario all emission from the pulsar will cease during 
nulls. The same might be true in the nulling model of Jones (\cite{PBJ1981}), 
which postulates a reduction in the surface electric potential on the polar cap, 
due to a reduction in the mean nuclear charge of surface ions. Of course what 
exactly happens in either models will depend crucially upon the specific 
assumptions made and the detailed physics incorporated; but checking for the 
presence (or absence) of nulling at higher energies, in pulsars that exhibit 
radio nulling, might be an important clue to understanding the emission 
mechanism of rotation powered pulsars. It would be ideal to observe such pulsars
simultaneously at radio and higher energies.

Now consider mode changing. According to Filippenko \& Radhakrishnan (\cite{FR1982})
this is merely radio nulling at ``certain regions of the polar gaps''. Therefore in 
their model mode changing might not occur at all at higher energies, which can be 
verified by the current and similar works.

The x-ray data were obtained by the Proportional Counter Array (PCA) aboard the 
RXTE x-ray observatory (ObsId numbers 10203-01-01-00 to 10203-01-03-01). They 
consist of 23 data files observed during August/September 1996, in the EVENT 
mode, combining photons from all five Proportional Counter Units, and also from 
both halves of all three Xenon anode layers of each PCU. Channels 50 to 249 of 
the PCA were combined, corresponding to the energy range 13.3 to 58.4 keV. The 
initial data analysis used the FTOOLS software, while the latter part used 
self-developed software; this is explained in detail in Vivekanand \cite{MV2001}.

\begin{figure}
\resizebox{\hsize}{!}{\includegraphics{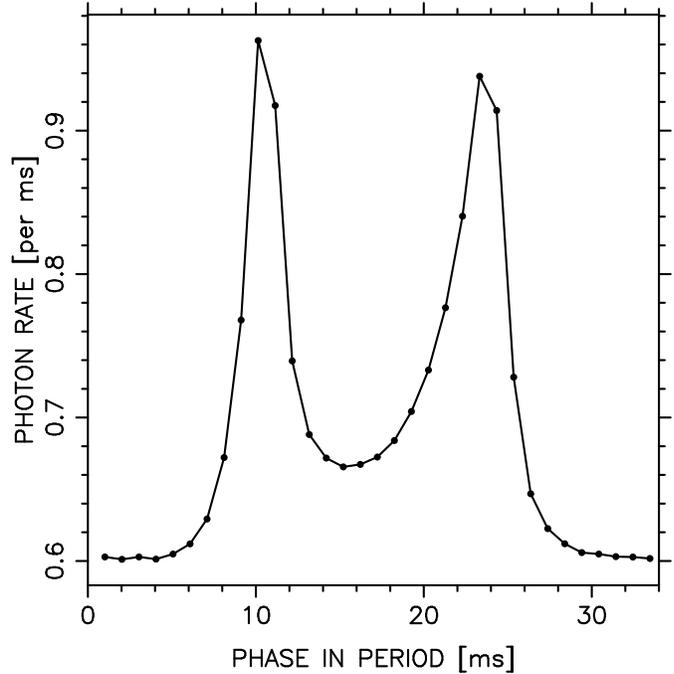}}
\caption{
	 Integrated profile of Crab pulsar after summing 1\,868\,112 periods 
         from 23 data files of RXTE. The abscissa is time (also called phase) 
	 within the period (in ms), while the ordinate is the average number 
	 of photons obtained in one time sample (1.013967 ms; this is called
	 ``synthesized'' time sample in Vivekanand \cite{MV2001}).
	}
\label{fig1}
\end{figure}

Fig.~\ref{fig1} shows the integrated profile of Crab pulsar for 1\,868\,112 
periods. Samples 5 to 30 are considered to represent the {\it on-pulse} window, 
and the rest of the seven samples the {\it off-pulse} window, although the Crab 
pulsar might emit x-rays all through its period. Details of the analysis in the 
coming sections can be found in Vivekanand \cite{MV1995}, Vivekanand \& Joshi 
\cite{VJ97}, Vivekanand et al. \cite{VAM1998}, Vivekanand \cite{MV2000} and 
Vivekanand \cite{MV2001}; they will be described only briefly here.

\section{Pulse nulling}

The coherent microwave radio radiation of some rotation powered pulsars often 
ceases for durations ranging from one to several hundred periods; this is 
known as ``pulse nulling'' (Backer \cite{DB70}; Ritchings \cite{RT76}). During
nulls the radio emission decreases by at least three orders of magnitude 
(Vivekanand \& Joshi \cite{VJ97}).

\begin{figure}
\resizebox{\hsize}{!}{\includegraphics{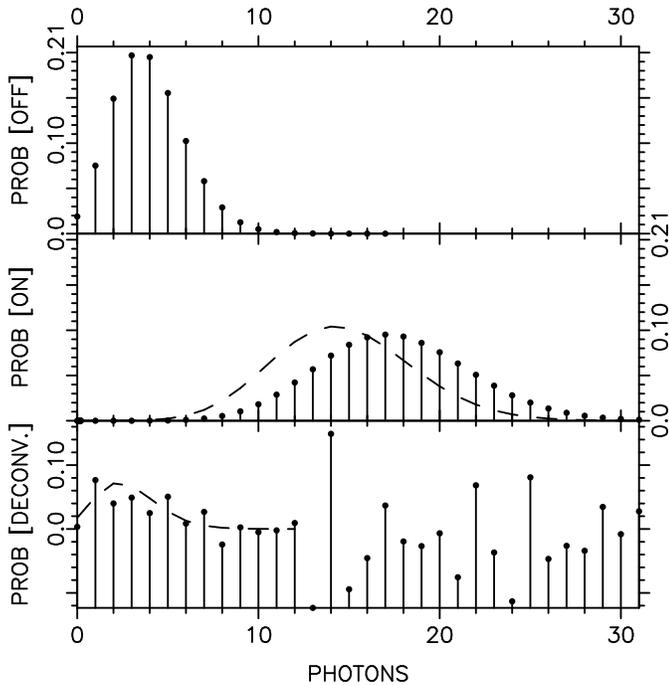}}
\caption{
	 Abscissa is the number of photons (observed or expected) while the 
	 ordinate is the probability of obtaining those photons. The dots in
	 the top two panels refer to observations in the off-pulse (7 samples) 
	 and on-pulse (26 samples) windows, respectively. Dead time correction 
	 for the PCA spreads the probability over $\approx$ 6\% of the abscissa 
	 around integer values, which has been corrected for. The dashed curve 
	 in the second panel is the expected (normalized) probability for 
	 nulled periods, if at all they exist, based on the off-pulse 
	 probability in the first panel. The third panel shows the result of 
	 deconvolving the observed and expected probabilities in the second panel; 
	 the dashed curve is the expected result in the absence of nulling.
	}
\label{fig2}
\end{figure}

Fig.~\ref{fig2} shows the observed probability of number of photons in the 
off-pulse and on-pulse windows for $\approx$ 1\,868\,112 periods. The off- and 
on- window data fit to Poisson distributions with mean values of 3.97 and 17.57 
photons, respectively; the corresponding $\chi^2$ are 12.3 and 38.2, for 18 
and 40 degrees of freedom, respectively, which implies that the fits are good.

Any nulled periods in the data will have a Poisson distribution of photons in 
the on-pulse window with a mean value of 3.97 $\times$ 26 / 7 = 14.75 photons, 
which is plotted as the dashed curve in the second panel of Fig.~\ref{fig2}. 
The observed on-pulse window probability in the second panel (dots) will be the 
convolution of the nulled distribution (dashed curve) with the true non-nulled 
distribution, the relative weights of the two distributions depending upon the 
fraction of nulled periods (Ritchings \cite{RT76}; Vivekanand \cite{MV1995}). 
The third panel of Fig.~\ref{fig2} shows the result of deconvolving the two 
probability distributions in the second panel. One expects the result to 
contain (in general) two distributions: (a) a Poisson distribution with a mean 
value of 17.57 $-$ 14.75 = 2.82 photons, which is plotted as the dashed curve 
in the third panel of Fig.~\ref{fig2}; this represents the true, non-nulled 
photon distribution of the Crab pulsar in the on-pulse window, and (b) a Dirac 
delta function at zero photons, which represents the nulled periods, if at all
(Ritchings \cite{RT76}; Vivekanand \cite{MV1995}).

\begin{figure}
\resizebox{\hsize}{!}{\includegraphics{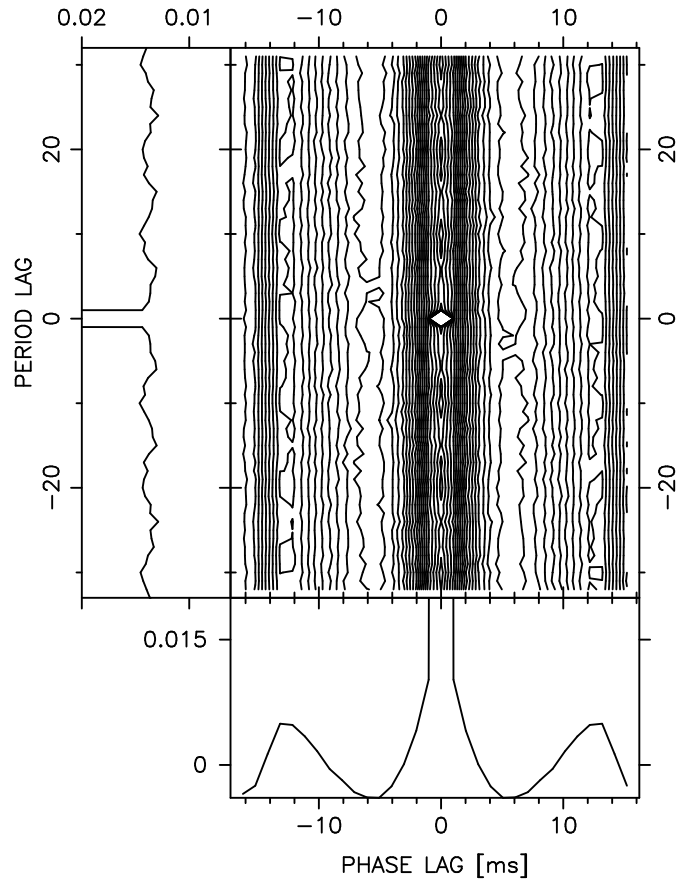}}
\caption{
	 Two dimensional autocorrelation of Crab pulsar x-ray flux as a function 
	 of phase and period lags, for 1\,868\,112 periods. Segments of two 
	 dimensional data, consisting x-ray flux at 32 time samples in 16\,384 
	 periods, were used; the details are described in Vivekanand et al. 
	 (\cite{VAM1998}). The peak autocorrelation was normalized to 1.0 in each 
	 segment before averaging. All 16 phase lags but only 32 period lags 
	 (both $\pm$ lags) have been plotted. The bottom panel is a cut at zero 
	 period lag; the secondary peaks are due to the two peaks of the 
	 integrated profile in fig.~\ref{fig1}. The left panel is a cut at zero
	 phase lag. The weak quasi periodic feature is due to folding the pulsar 
	 data at a period that is not an integral multiple of the original 
	 sampling interval, which is 265 times the basic time resolution of the 
	 data (Vivekanand \cite{MV2001}). The synthesized sampling interval 
	 (each period has 33 such samples) differs from the former by 1.013967 
	 $-$ 1.010895 = 0.003072 ms, which causes a difference of 0.003072 
	 $\times$ 33 / 1.013967 = 0.09998 synthesized samples per period; this 
	 causes a periodicity of 1.0 / 0.09998 $\approx$ 10 periods in the data. 
	 This was confirmed by obtaining light curves with a slightly different 
	 sampling interval (1.0 ms), and folding the data as described in 
	 Vivekanand et al. \cite{VAM1998} (note that this method is only suitable 
	 for radio data, since it does not preserve the Poisson nature of photon 
	 statistics). The quasi periodic spectral feature almost disappeared.
	}
\label{fig3}
\end{figure}

Now it is well known that the deconvolution process is very sensitive to errors
in the data. Clearly the deconvolution has not worked properly in Fig.~\ref{fig2}, 
in spite of reducing the number of bins in the distribution, 
attempting a Wiener filter, etc. More data is needed for a robust result. 
However, the deconvolved distribution is not inconsistent with the dashed curve
up to 12.0 photons along the abscissa; and there is no evidence for an excess 
probability (over the dashed curve) at 0.0 photons. Therefore one concludes that 
Crab pulsar does not display the pulse nulling phenomenon at x-ray energies. 
However one needs more data or better techniques for a quantitative result.

For the record, the deconvolution process had not worked either for PSR 
J0437-4715 in Fig.~2 of Vivekanand et al. (\cite{VAM1998}); the lower panel of 
their Fig.~2 shows the original distribution only. This explanation was left 
out inadvertently in that paper. However the nulling results of Vivekanand et al. 
(\cite{VAM1998}) do not change, because they were verified with a much coarser 
binning, where the deconvolution worked. Moreover, the deconvolution is not 
expected to significantly modify their original distribution, since the 
deconvolving function (off-pulse distribution) is about six times narrower than 
the function deconvolved (on-pulse distribution). In Fig.~\ref{fig2} above, the
deconvolving function is comparable in width to the function deconvolved, which 
might be one of the reasons why the deconvolution is not working, in spite of
the factor $\approx$ 10 or more amount of data available here.

\section{Systematic sub pulse drifting}

In some rotation powered pulsars, the radio emission is typically confined to 
two sub pulses, which march across the integrated profile with each successive 
period. This pattern repeats periodically (see Lyne \& Ashworth \cite{LA83}, 
and Vivekanand \& Joshi \cite{VJ97}, for two of the best examples). These are
known as ``drifting sub pulses''.

There are two standard methods of checking for drifting sub pulses. One is to 
plot the pulsar flux as a two dimensional function of the phase within the period 
and the period number, and obtain the two dimensional autocorrelation. Fig.~\ref{fig3} 
shows the average two dimensional autocorrelation of Crab pulsar at
x-rays; one does not notice here the characteristic sloping bands due to drifting 
sub pulses.

\begin{figure}
\resizebox{\hsize}{!}{\includegraphics{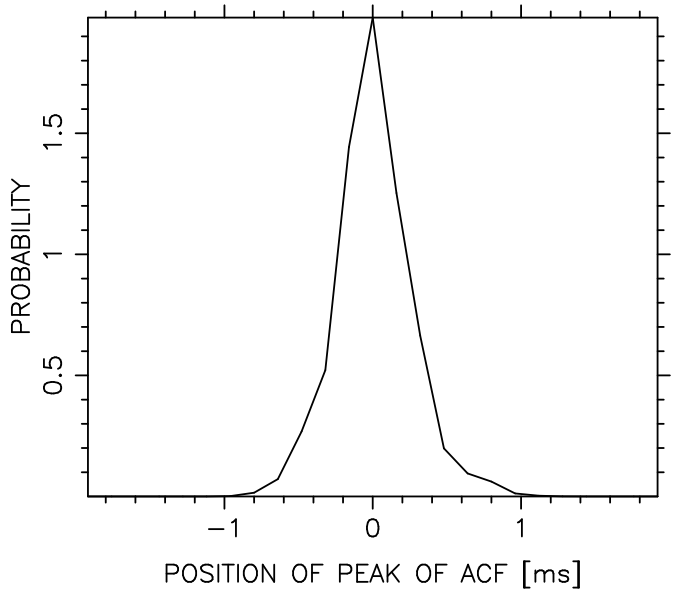}}
\caption{Probability density of occurrence of the peak in a horizontal cut 
	 at +1 period lag in Fig.~\ref{fig3}. Each peak is represented by 
	 a Gaussian of unit area, centered at the position of the peak,
	 of rms width equal to the nominal error on the position. There
	 are 25 bins along the abscissa in the range $\pm$ 2.027934 ms (2 
	 synthesized time samples). Only those cuts with positional errors
	 $\le$ 0.2 ms were included.
	}
\label{fig4}
\end{figure}

The second method (Taylor et al. \cite{TMH75}) is to take cuts in Fig.~\ref{fig3} at 
period lag +1, and plot a distribution of the positions of the 
peaks of the autocorrelation. The sub pulse drifting phenomenon shows up in 
this plot as an off centered distribution, depending upon the drift rate. 
Fig.~\ref{fig4} shows that probability density for 115 such cuts (see Fig.~4 of 
Vivekanand et al. \cite{VAM1998} for the corresponding plot for PSR 
J0437$-$4715 at radio wavelengths). The distribution in Fig.~\ref{fig4} is 
quite symmetric; its mean value lies at sample 0.007$\pm$0.113 ms, which is 
consistent with there being no systematic drifting sub pulses in the Crab pulsar 
at x-rays.

\section{Stability of integrated profile}

The integrated profiles of rotation powered pulsars at radio wavelengths are 
generally very stable, after folding $\approx$ 100 periods. They are supposed 
to be as unique to pulsars as finger prints are to humans. However in some 
of them, the integrated profile changes to another stable shape, remains in 
that shape for $\approx$ 100 periods, before reverting back to the original 
integrated profile.  This is known as ``mode changing'' (see Manchester \& 
Taylor \cite{MT1977}). This should not be confused with the slower variations
of integrated profiles seen in ms pulsars (Vivekanad et al. \cite{VAM1998}; 
Kramer et al. \cite{MK1999}).

\begin{figure}
\resizebox{\hsize}{!}{\includegraphics{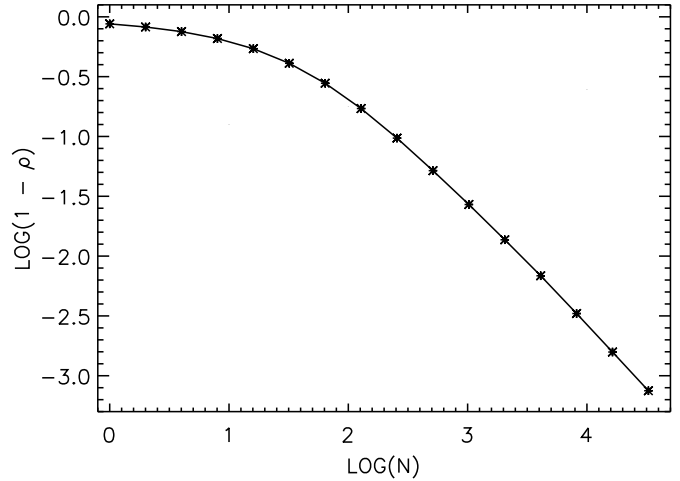}}
\caption{
	 Normalized correlation coefficient $\rho$ between the master integrated
	 profile, consisting of all 1\,868\,112 periods, and sub integrated 
	 profiles, consisting of N periods, both plotted in units of logarithm 
	 to the base ten. At each N there are 1\,868\,112 / N $\rho$s that have
	 been averaged. No correction has been done for background photon counts,
	 unlike in Fig.~9 of Vivekanand et al. \cite{VAM1998}, which has been
	 corrected for receiver noise); however this is not expected to introduce
	 kinks in the above figure.
	 }
\label{fig5}
\end{figure}

This is tested by correlating the master integrated profile, formed by folding
the entire data, with sub integrated profiles consisting of N periods, for 
various values of N. Fig.~\ref{fig5} shows the plot of the logarithm of $1 - 
\rho$, where $\rho$ is the average correlation coefficient, as a function of 
logarithm of N, with N differing by powers of 2 (Helfand et al. \cite{HMT75}; see 
Fig.~9 of Vivekanand et al. \cite{VAM1998} for the corresponding plot for PSR 
J0437$-$4715 at radio wavelengths). The correlation $\rho$ is very low for small 
N, since the x-ray emission of Crab pulsar is dominated by photon noise. 

This is in contrast to radio pulsars, where even the individual periods bear some 
resemblance to the integrated profile. Table 2 shows that, at x-rays the integrated 
profile of Crab pulsar requires $\approx$ 100 times more periods to stabilize, in 
comparison to PSR J0437$-$4715 at radio wavelengths.  For N larger than $\approx$ 500, 
$\rho$ shows no change of slope in Fig.~\ref{fig5}, which would have been the 
signature of some perturbing influence on the stability of the integrated profile on 
those time scales (for example, mode changing). This trend continues for N 
$\approx$ 32\,000 periods. One can therefore conclude that the Crab pulsar does not 
display the mode changing phenomenon at x-rays.

\begin{table}[h]
\begin{tabular}[t]{ccc}
\hline
\hline
LOG(1 - $\rho$) & \multicolumn{2}{c}{LOG(N)} \\
\hline
\hline
 & PSR0531 & PSR0437 \\
\cline{2-3}
\cline{2-3}
-1.0  &   2.4 	& 0.3 \\
-2.0  &   3.5 	& 1.3 \\
-3.0  &   4.4 	& 2.5 \\
\hline
\end{tabular}
\caption{Comparison of LOG(N) values from Fig.~\ref{fig5} for Crab pulsar at 
	 x-rays, and from Fig.~9 of Vivekanand et al. \cite{VAM1998} for PSR 
	 J0437$-$4715 at radio wavelengths.}
\end{table}

\section{Discussion}

The Crab pulsar does not display nulling, systematic sub pulse drifting or mode changing 
at x-ray energies. Its radio behavior has not been studied so far (see Manchester \& 
Taylor \cite{MT1977}). The difficulty is that Crab pulsar observations at radio wavelengths 
using single dishes are flooded by the continuum emission from the Crab nebula; it is 
difficult to obtain high signal to noise ratio observations of single pulses. Crab pulsar 
observations at radio wavelengths using interferometers will overcome this problem, but 
these are technically difficult observations, and no results have been published until now
(Hankins \cite{TH2001}).

So far, none of the above three phenomena have found a satisfactory explanation,
and they have remained essentially radio wavelength phenomena; but it is possible 
that similar observations at higher energies might provide the clues for resolving 
them.

Consider nulling first. While discussing the theoretical problems in understanding
this phenomenon at radio wavelengths, Michel \cite{FCM1991} writes ``Rather than 
the micro physics faltering (e.g. bunching) the global physics might falter (e.g. 
current flow interruption), ...'' on page 68. Now, it is likely that it is the 
latter mechanism that might manifest itself in the high energy (e.g. x-ray) 
observations than the former. Therefore finding a rotation powered pulsar that 
nulls either (1) at both radio and high energies, or (2) at one and not the other 
energy, might provide an important clue to understanding the phenomenon.

Consider next drifting subpulses at radio wavelengths. The Ruderman and Sutherland 
\cite{RS} model invokes an electric discharge that drifts systematically in the 
crossed electric and magnetic fields on the polar cap. Thus drifting is supposed 
to be essentially an inner-gap phenomenon; it is not clear that the electric and 
magnetic fields are suitable in the outer-gap for the observed drifts to occur. 
Now, finding a rotation powered pulsar that shows systematic subpulse drifting at 
high energies also might be a strong constraint for the outer-gap models.

Finally, it is possible that both inner and outer gaps can not exist simultaneously 
in the same rotation powered pulsar; operation of one might extinguish the other 
(CHR). In that case, simultaneous radio and high energy observations of rotation 
powered pulsars might be fruitful. For example, if one notices that the radio and 
x-ray nulls (if at all) are mutually exclusive, then one would have important 
information concerning the possible interplay between the two gaps, in terms of 
electric currents from one gap shorting out the electric fields of the other gap.

The implications of this work depend upon the radio behavior of the Crab pulsar, 
where none, or some or all of the above three phenomena might be found. If none
of these are found in the radio, then the current results are consistent with 
the Crab pulsar emitting the radio and x-rays from the outer gap alone (CHR),
although it would not be considered a conclusive proof.

However, suppose 
the Crab pulsar shows the pulse nulling phenomenon in the radio; then its absence 
at x-rays implies that loss of coherence, of some other such physics is responsible 
for nulling, while the basic relativistic charges from the Crab pulsar continue 
to emit the higher energies. In other words it is the micro physics faltering 
rather than the global physics. An interesting possibility is if the Crab pulsar
shows nulling for a very small fraction of the time at both the radio as well as 
at the x-rays, which might have been impossible to detect by this work, then
the global physics would be held responsible for nulling. A much more interesting 
possibility would be if nulling is not simultaneous at the two wavelengths; this
might necessitate the operation of both the gaps (inner and outer) in the Crab 
pulsar, as discussed above.

Suppose the Crab pulsar displays systematic drifting subpulses. Then their absence 
at x-rays might be strong evidence for the two emissions arising in the two 
different gaps. In other words the radio and x-ray emissions are probably highly 
decoupled from each other.

\begin{acknowledgements}
I thank the referee for very useful corrections and suggestions.
This research has made use of (a) High Energy Astrophysics Science Archive 
Research Center's (HEASARC) facilities such as their public data archive, and 
their FTOOLS software, and (b) NASA's Astrophysics Data System (ADS) 
Bibliographic Services. I thank them for their excellent services.

\end{acknowledgements}

\end{document}